\documentclass[conference]{IEEEtran}
\IEEEoverridecommandlockouts

\usepackage{cite}
\usepackage{amsmath,amssymb,amsfonts}
\usepackage{amsthm}
\usepackage{algorithmic}
\usepackage{graphicx}
\graphicspath{{Figures/}}   
\usepackage{textcomp}
\usepackage{xcolor}
\usepackage{booktabs}
\usepackage{tikz}
\usetikzlibrary{arrows.meta,positioning,calc,fit,backgrounds}
\def\BibTeX{{\rm B\kern-.05em{\sc i\kern-.025em b}\kern-.08em
    T\kern-.1667em\lower.7ex\hbox{E}\kern-.125emX}}

\newtheorem{proposition}{Proposition}
\newtheorem{theorem}{Theorem}

\newtheorem{assumption}{Assumption}
\newtheorem{remark}{Remark}

\newcommand{\Dage}{\Delta}
\newcommand{\Dec}{\Delta_{R}}
\newcommand{\Dk}{\Delta_{(k)}}
\newcommand{\E}{\mathbb{E}}
\newcommand{\PR}{\mathbb{P}}
\newif\ifshowtodos
\showtodostrue

\ifshowtodos
  \newcommand{\TODO}[1]{\textcolor{red}{\textbf{[TODO: #1]}}}
  \newcommand{\CITE}[1]{\textcolor{red}{\textbf{[CITE: #1]}}}
\else
  \newcommand{\TODO}[1]{}
  \newcommand{\CITE}[1]{}
\fi

\definecolor{cSens} {HTML}{3F6EA5}   
\definecolor{cSensF}{HTML}{E7EEF7}   
\definecolor{cFus}  {HTML}{2F6DB5}   
\definecolor{cFusF} {HTML}{DCE8F5}
\definecolor{cRule} {HTML}{C8603F}   
\definecolor{cRuleF}{HTML}{FBE6DC}
\definecolor{cOps}  {HTML}{5B9B4F}   
\definecolor{cOpsF} {HTML}{E4F1DF}
\definecolor{cAcc}  {HTML}{C9972F}   
\definecolor{cMut}  {HTML}{7A8292}   
\definecolor{cWash} {HTML}{F4F5F7}   

\tikzset{
  pics/icradar/.style={code={
    \draw (-1.6mm,-1.4mm) -- (1.6mm,-1.4mm);
    \draw (0,-1.4mm) -- (0,-0.4mm);
    \draw (-1.45mm,-0.4mm) arc (180:0:1.45mm);
    \draw (0.75mm,0.95mm) arc (-55:55:0.7mm);
    \draw (1.55mm,0.95mm) arc (-55:55:1.1mm);
  }},
  pics/iccam/.style={code={
    \draw[rounded corners=0.4pt] (-1.9mm,-1.3mm) rectangle (1.9mm,1.0mm);
    \draw (-0.65mm,1.0mm) -- (-0.4mm,1.55mm) -- (0.5mm,1.55mm) -- (0.75mm,1.0mm);
    \draw (0,-0.15mm) circle (0.7mm);
  }},
  pics/icant/.style={code={
    \draw (0,-1.5mm) -- (0,0.95mm);
    \draw (-1.0mm,-1.5mm) -- (0,-0.2mm) -- (1.0mm,-1.5mm);
    \draw[fill] (0,1.2mm) circle (0.25mm);
    \draw (0.55mm,0.7mm) arc (-45:45:0.75mm);
    \draw (-0.55mm,0.7mm) arc (225:135:0.75mm);
  }},
  pics/icsat/.style={code={
    \draw[rounded corners=0.3pt] (-0.65mm,-0.9mm) rectangle (0.65mm,0.9mm);
    \draw (-2.1mm,-0.6mm) rectangle (-0.8mm,0.6mm);
    \draw (2.1mm,-0.6mm) rectangle (0.8mm,0.6mm);
    \draw (0,0.9mm) -- (0.45mm,1.6mm);
  }},
}

\newcommand{\ResRatioMin}{2.4}
\newcommand{\ResRatioMax}{5.2}
\newcommand{\ResTauC}{6.49}
\newcommand{\ResMaxCdfError}{0.0019}
\newcommand{\ResMaxMeanError}{0.0034}
\newcommand{\ResNDecisions}{1\,196\,724}
\newcommand{\ResThmVsModel}{4.5e-07}
\newcommand{\ResThmVsSim}{0.045}
\newcommand{\ResAgeKone}{0.75}
\newcommand{\ResAgeKsix}{7.10}
\newcommand{\ResAgeRatio}{9.5}
\newcommand{\ResViolKfive}{14}
\newcommand{\ResViolKsix}{62}
\newcommand{\ResStaleFrac}{45}
\newcommand{\ResNFreeSources}{2}
\newcommand{\ResDecTotal}{4.22}
\newcommand{\ResShareSampling}{45}
\newcommand{\ResShareDelay}{13}
\newcommand{\ResShareCorrob}{42}
\newcommand{\ResCorrobOverDelay}{3.2}
\newcommand{\ResShareCorrobTenX}{34}
\newcommand{\ResShareCorrobHomog}{34}
\newcommand{\ResShareCorrobKthree}{27}
\newcommand{\ResDomTopName}{UAS feed}
\newcommand{\ResDomTop}{42}
\newcommand{\ResDomTwo}{70}
\newcommand{\ResDomCameras}{2.1}
\newcommand{\ResPoissonError}{0.051}
\newcommand{\ResPeriodicEight}{1.78}
\newcommand{\ResPoissonEight}{2.72}
\newcommand{\ResLossWorstName}{EO camera}
\newcommand{\ResLossWorstAge}{7.1}
\newcommand{\ResLossWorstViol}{62}
\newcommand{\ResLossBestName}{AIS}
\newcommand{\ResLossBestAge}{5.2}
\newcommand{\ResLossBestViol}{31}
\newcommand{\ResAvailMeanLo}{4.21}
\newcommand{\ResAvailMeanHi}{4.81}
\newcommand{\ResAvailTailLo}{8.7}
\newcommand{\ResAvailTailHi}{10.6}
\newcommand{\ResAvailViolLo}{14}
\newcommand{\ResAvailViolHi}{24}
\newcommand{\ResCommErrZero}{0.093}
\newcommand{\ResCommErrSpread}{0.001}

\begin{document}

\title{Decision Age: Information Freshness under Fusion Rules in Multi-Sensor Command and Control}

\author{Victor Monzon Baeza \\
\IEEEauthorblockA{\textit{Universidad Internacional de La Rioja (UNIR)}, Logro\~{n}o, Spain \\
victor.monzon@unir.net \\
ORCID: 0000-0002-0035-3944}
}

\maketitle

\begin{abstract}
Command and control systems decide automatically on evidence that a fusion rule requires from several sensors. The Age of Information measures how current a datum is when it reaches the fusion node, but existing multi-source results treat additional sources as redundancy: the age at the monitor falls as sources are added. This paper addresses the opposite regime. When a rule demands corroboration the decision waits for the slowest required source, so the binding quantity is a maximum, not a minimum, of source ages. We define the decision age of an arbitrary monotone fusion rule, obtain its distribution in closed form for periodic and event-driven sources, and show that the mean decision age grows linearly with the corroboration level, from the floor of the age sawtooth to its peak. We add a dominance criterion that identifies which sensor governs decision freshness, and a measurement methodology for fielded systems. Freshness becomes a verifiable requirement, and adding sensors is shown to make automatic decisions older rather than fresher.
\end{abstract}


\section{Introduction}\label{sec:intro-motivation}

A command and control (C2) system, and equally an intelligence, surveillance and reconnaissance (ISR) chain terminating in an automatic alarm, is a distributed state-estimation and decision machine: sensors observe, a fusion process associates and tracks, a decision rule fires, and an operator or an effector acts. Such systems are specified, procured and accepted through metrics of the means, as in the recent system-level design of a nationwide microwave backbone for air and maritime surveillance \cite{baeza2026backbone}, validated through link margin, traffic segregation and end-to-end availability. Each such quantity is necessary, and each is a property of the infrastructure; none measures the \emph{product}, which is how current the knowledge is at the instant the decision is taken. A requirement of the form ``link latency below a given bound'' prescribes a solution before the problem has been stated and, as we show, can be met in full by a system whose decisions are chronically stale.

The Age of Information (AoI) captures exactly that currency \cite{kaul2012,yates2021}, and now covers optimal sampling \cite{sun2017}, scheduling \cite{kadota2019} and freshness in tactical networks, where it has been used both to choose the queueing discipline of a deployed military sensor mesh \cite{chache2022} and to characterise how quickly information propagates through a tactical topology \cite{srivastava2023,srivastava2026}. It characterises, however, the age of the data \emph{at the monitor}, and its multi-source extensions all treat additional sources as \emph{redundancy}: whether they report distinct processes through a shared server \cite{yates2019}, carry correlated observations \cite{tripathi2022}, or simply duplicate an observation so that the monitor keeps whichever copy arrives first, an update from any one of them lowers the age. The structure of the network behaves the same way: increasing connectivity lowers the age at every node, from a ring to a fully connected graph \cite{srivastava2026}. Within that body of work, more sources and more connectivity mean fresher information.

\begin{figure}[t]
\centerline{\includegraphics[width=\columnwidth]{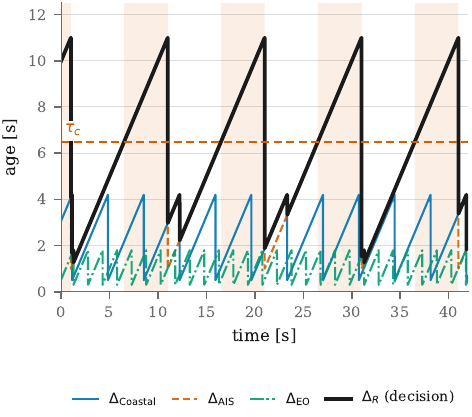}}
\caption{The effect, before any of the analysis. Three sources report at
different cadences; each age rises at one second per second and falls on
delivery, never to zero but to the chain delay of the observation that
arrived. The heavy trace is the age of a decision that requires all three.
It does not follow the average of the sources, nor the freshest: it follows
the slowest one that the rule requires. The decision is older than the
operational tolerance $\tau_c$ in the shaded intervals, \ResStaleFrac{}\% of
the time, while the two fast sensors are never stale themselves and
contribute nothing to the freshness of the decision. Improving either of them
cannot change that number. Section \ref{sec:results} returns to this figure
with the scenario and the numbers behind it.}
\label{fig:sawtooth}
\end{figure}

C2 decisions do not work that way. An automatic alarm, an engagement authorisation or a track promotion is taken on evidence that a \emph{fusion rule} requires from several heterogeneous sources: radar corroborated by AIS, a detection confirmed by imagery, a classification supported by two independent sensing modalities. The decision cannot then be taken until the \emph{slowest} required source has reported, so the binding quantity is a \textbf{maximum} of source ages rather than a minimum, and the inherited intuition is reversed: a corroborating source improves detection and \emph{degrades} the freshness of the decision. Fig. \ref{fig:sawtooth} shows the effect on three sources of the surveillance suite used later in the paper: the decision follows the slowest source its rule requires, and the two fast sensors, never stale themselves, contribute nothing to its freshness. The rest of the paper is an attempt to say that precisely, and to say what follows from it for the design of a C2 system. The point is a practical one, because multi-domain programmes keep adding sensors while the infrastructure that carries them is reshaped by virtualisation, artificial intelligence and emerging sixth-generation technologies \cite{baeza2024tactical}, both on the assumption that more sources and more capacity yield a better picture. Connectivity, then, is not the only structure that governs freshness: the decision rule is another, and it acts in the opposite direction. Our claim is correspondingly narrow: not that freshness has been overlooked in military systems, but that the ages of several sources have never been \emph{composed} through the decision rule that consumes them.

\subsection{Contributions}

\begin{enumerate}
\item We define the \textbf{decision age} of an arbitrary monotone fusion rule and obtain its distribution in closed form for periodic and event-driven sources (Sections \ref{sec:model} and \ref{sec:analysis}).

\item We characterise its growth with the corroboration level, and show that the decision rule alone drives it from the floor of the age sawtooth to its peak.

\item We give a \textbf{dominance criterion} identifying which sensor governs decision freshness, and hence where engineering effort changes the outcome and where it does not.

\item We give a \textbf{measurement methodology} for fielded C2 systems (Section \ref{sec:measure}): an exact event-driven age estimator, a clock synchronisation budget, and query-instant sampling.
\end{enumerate}

The analysis is instantiated on a port-approach surveillance scenario built from publicly documented sensor parameters, though the formulation is generic to any multi-source C2 or ISR chain. All results are validated against discrete-event simulation, and the simulator and the parameter set are released.

\section{Related Work}\label{sec:related}

\subsection{Age of Information}

AoI was introduced in \cite{kaul2012} with the now-classical result that, for an M/M/1 status-update system, there exists an interior optimal generation rate: updating too slowly leaves the monitor stale between samples, while updating too quickly fills the queue and delivers stale packets. The survey in \cite{yates2021} collects the state of the art, and \cite{kaswan2025} does the same for the multi-node case. Two strands are relevant here. The first is optimal sampling, where \cite{sun2017} shows that the zero-wait policy is not in general optimal. The second is minimum-age scheduling, where index policies with constant-factor guarantees are available \cite{kadota2019}.

\subsection{What ``Multiple Sources'' Means in the AoI Literature}

The term \emph{multiple sources} carries a specific meaning in this literature that differs from the setting considered here, and the distinction is the premise of the paper.

In \cite{yates2019} several sources report \emph{distinct} processes through a shared server, and the quantity derived is the age \emph{of each source} at the monitor. The ages are never combined, because they describe different states; the coupling between sources is the queue, not the information. A second family has several sensors observe \emph{the same} process in parallel, with the monitor retaining whichever observation arrives first; the age at the monitor is then by construction a \emph{minimum} over the sources, and decreases as sources are added. A third models correlation at the sensing level, where an update from one source also refreshes its neighbours with some probability, which again lowers ages \cite{tripathi2022}. A fourth is structural rather than statistical: in gossip networks the age at a node falls as the connectivity of the topology rises, from $O(\sqrt{n})$ on a ring to $O(\log n)$ on a fully connected graph, and the intermediate structures, including generalised rings, grids and hypercubes, have recently been characterised \cite{srivastava2023,srivastava2026}. Applied work in tactical sensor networks follows the same logic, choosing the queueing discipline that keeps the age at the gateway low \cite{chache2022}. A fifth family admits that the update itself takes time to produce, modelling transmission and computation as tandem queues and deriving the age and peak age in closed form \cite{qin2023}. That family comes closest to the setting of this paper, and is explicit about the step it takes: the sensors of a device ``are usually fused together to generate a perception task'', after which the task generation process is abstracted as Poisson \cite{qin2023}. The fusion is named and then removed from the model. It is that step which is not taken here.

The redundancy reading is not universal, and one exception is worth stating because it is the closest existing work to the object defined here. Where a fused product waits for every sensor that contributes to it, its delay is set by the slowest of them, and collaborative perception systems are modelled in exactly that way: the delay of a region is a maximum over its sensors, and the age is reset to that delay \cite{zhu2026}. The maximum therefore already appears in the literature. What does not appear is the maximum as a \emph{variable}: there the contributing set is fixed and complete, the maximum enters a scheduling objective through its mean, no distribution is derived for it, and timeliness is an empirically fitted penalty rather than a requirement.

Table \ref{tab:related} places these families side by side with the present paper. Read along its last two columns, the pattern is uniform: wherever the metric is elapsed time, the sources either are not combined at all or combine as a minimum, and the age falls as sources or links are added. The only entries in which an order statistic appears, or in which the age rises with the number of components, use version age and carry no decision semantics.

The present work addresses the complementary regime. The $n$ sources observe the same physical state, as in the second family, but the decision rule \emph{requires} several of them, so the binding quantity is a maximum rather than a minimum, and the age increases as the corroboration requirement is tightened.

\begin{remark}
Across these families the age at the monitor is a minimum, a per-source quantity, or a mixture that decreases with the number of sources or links; the one setting in which a maximum arises \cite{zhu2026} takes the contributing set as fixed and complete. The contribution claimed here is therefore not the maximum itself, but three things that follow once the contributing set is governed by a rule: that the rule is a \emph{design variable}, so the binding quantity sweeps from a minimum to a maximum as $k$ runs from $1$ to $n$; that the resulting age has a distribution in closed form rather than a mean used as a plug-in; and that this distribution supports a probabilistic operational requirement rather than a fitted penalty. The recent survey of the multi-node literature \cite{kaswan2025} enumerates the axes along which the problem has been generalised, namely topology, resource constraints, adversarial models, source reliability and information mutation; composition through a decision rule is not among them.
\end{remark}

\begin{table*}[t]
\caption{Positioning with respect to the multi-source and multi-node AoI literature}
\begin{center}
\footnotesize
\renewcommand{\arraystretch}{1.2}
\begin{tabular}{@{}lllll@{}}
\toprule
\textbf{Work} & \textbf{Sources observe} & \textbf{Ages combine by} & \textbf{Metric} & \textbf{More sources or links} \\
\midrule
Kaul, Sun \emph{et al.} \cite{kaul2012,sun2017}
  & one source, one process    & not applicable          & AoI (elapsed)      & not applicable \\
Yates and Kaul \cite{yates2019}
  & distinct processes         & not combined            & AoI (elapsed)      & per-source age rises (queue) \\
Tripathi and Modiano \cite{tripathi2022}
  & correlated states          & a minimum               & AoI (elapsed)      & age falls \\
Chache \emph{et al.} \cite{chache2022}
  & a tactical sensor mesh     & queueing discipline     & AoI (elapsed)      & age falls \\
Srivastava and Ulukus \cite{srivastava2023,srivastava2026}
  & one process, gossiped      & network topology        & version age        & age falls as connectivity rises \\
Zhu \emph{et al.} \cite{zhu2026}
  & one scene, by region       & all-of-$n$, fixed set   & AoI (elapsed)      & delay set by the slowest sensor \\
Bayram \emph{et al.} \cite{bayram2024}
  & key shares of one object   & $k$-th order statistic  & version age        & age rises with $k$ \\
Shisher, Zhang \emph{et al.} \cite{shisher2022,zhang2025}
  & several modalities         & learned error function  & vector of ages     & model-dependent \\
Varshney \cite{varshney1997}
  & one state                  & $k$-of-$n$ fusion rule  & detection prob.    & detection improves \\
\midrule
\textbf{This paper}
  & \textbf{one state}         & \textbf{$k$-of-$n$ fusion rule} & \textbf{decision age (elapsed)} & \textbf{age rises with $k$} \\
\bottomrule
\end{tabular}
\label{tab:related}
\end{center}
\end{table*}

\subsection{Threshold Structures and Age}

Order statistics have appeared in age analysis in a different setting. In \cite{bayram2024} the age of a $k$-out-of-$n$ threshold structure is characterised on a gossip network, where the $n$ components are cryptographic key shares of a single object and $k$ is a decryption threshold. The order statistic there is taken over gossip edge activation times, the metric is version age rather than elapsed time, and no decision or alarm semantics are involved. The distinction between the two metrics is not cosmetic: version age counts how many updates a node is behind and does not grow while the source is idle, whereas the quantity an operational tolerance is written against grows in seconds whether or not anything is being transmitted \cite{kaswan2025}. The present work differs in all three respects: the $n$ components are independent observations of one physical state, the order statistic is taken over the stationary ages of sources with heterogeneous sampling periods, and $k$ is a corroboration requirement imposed by a decision rule, which is measured in elapsed time because the operational tolerance derives from threat kinematics.

\subsection{Freshness and Inference}

Closest in spirit is the remote-inference line of \cite{shisher2022,zhang2025}, which relates inference error to age and, in the multimodal case, to the \emph{vector} of per-modality ages. That work takes the error as a general, empirically characterised function of the age vector and solves a scheduling problem, namely which modality to transmit next. It does not compose the age vector through a decision rule, does not consider logical fusion rules, and yields no closed-form law for the age of the resulting decision.

\subsection{Decision Fusion}

Fusion rules requiring a fixed number of corroborating detectors are classical in distributed detection \cite{varshney1997}, where they are analysed on the detection and false-alarm axis. We are not aware of work that treats the same rules on the \emph{temporal} axis. The present paper crosses the two.

\section{System Model}\label{sec:model}

\subsection{Sensing Chains}

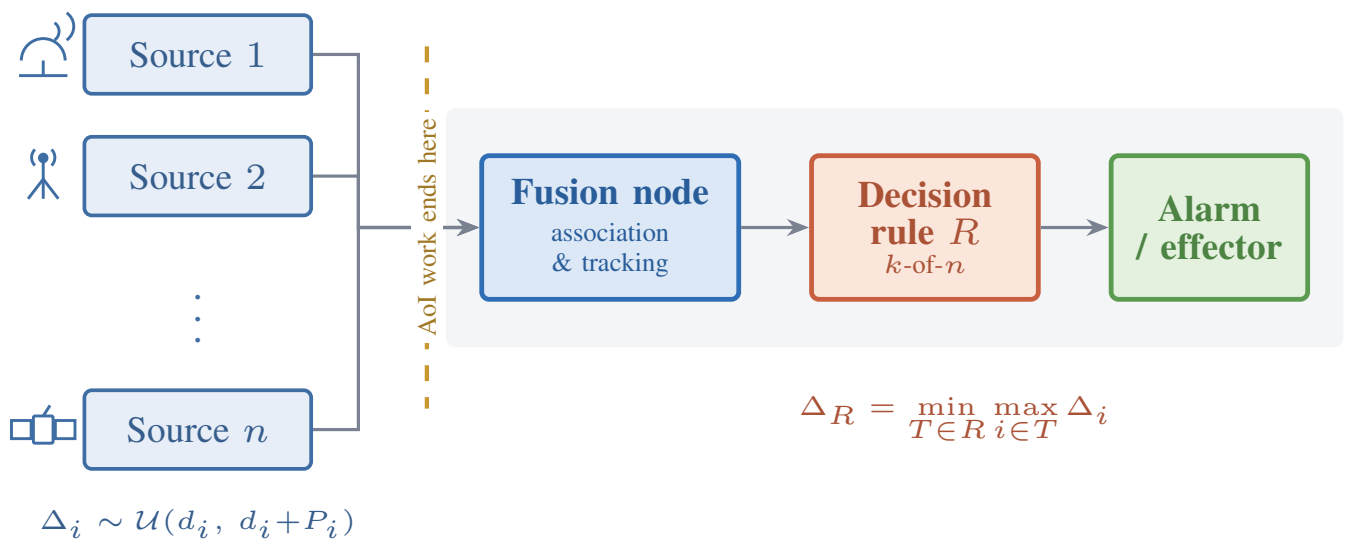
\begin{figure}[htbp]
\centerline{
\begin{tikzpicture}[font=\scriptsize,>={Stealth[length=4.2pt,width=3.2pt]},
  node distance=2.6mm,
  s/.style={fill=cSensF,draw=cSens,line width=.7pt,text=cSens!85!black,
            rounded corners=1.4pt,minimum width=15mm,minimum height=5.2mm,
            align=center,inner sep=1pt,font=\scriptsize},
  b/.style={fill=cFusF,draw=cFus,line width=.9pt,text=cFus!85!black,
            rounded corners=1.4pt,minimum width=17mm,minimum height=9.5mm,
            align=center,inner sep=1.5pt},
  r/.style={fill=cRuleF,draw=cRule,line width=.9pt,text=cRule!85!black,
            rounded corners=1.4pt,minimum width=15mm,minimum height=9.5mm,
            align=center,inner sep=1.5pt},
  o/.style={fill=cOpsF,draw=cOps,line width=.9pt,text=cOps!85!black,
            rounded corners=1.4pt,minimum width=13mm,minimum height=9.5mm,
            align=center,inner sep=1.5pt},
  ar/.style={->,draw=cMut,line width=.6pt}]

\node[s] (s1) at (5mm,0) {Source $1$};
\node[s,below=of s1] (s2) {Source $2$};
\node[below=1.8mm of s2,font=\footnotesize,text=cSens] (sd) {$\vdots$};
\node[s,below=1.8mm of sd] (sn) {Source $n$};

\pic[cSens,line width=.55pt] at ($(s1.west)+(-2.6mm,0)$) {icradar};
\pic[cSens,line width=.55pt] at ($(s2.west)+(-2.6mm,0)$) {icant};
\pic[cSens,line width=.55pt] at ($(sn.west)+(-2.6mm,0)$) {icsat};

\node[b,right=11mm of s2,yshift=-3.4mm] (fus)
     {\textbf{Fusion node}\\[-1pt]\tiny association\\[-2.5pt]\tiny \&\ tracking};
\node[r,right=4.5mm of fus] (rule)
     {\textbf{Decision}\\[-1pt]\textbf{rule} $R$\\[-2.5pt]\tiny $k$-of-$n$};
\node[o,right=4.5mm of rule] (op) {\textbf{Alarm}\\[-1pt]\textbf{/ effector}};

\begin{scope}[on background layer]
  \node[fill=cWash,rounded corners=1.8pt,
        fit=(fus)(rule)(op),inner xsep=2.2mm,inner ysep=3.0mm] (panel) {};
\end{scope}

\foreach \x in {s1,s2,sn} \draw[ar] (\x.east) -- ++(3.0mm,0) |- (fus.west);
\draw[ar] (fus) -- (rule);
\draw[ar] (rule) -- (op);

\draw[cAcc,line width=.8pt,dashed]
  ($(panel.north west)+(-1.2mm,4mm)$) -- ($(panel.south west)+(-1.2mm,-4mm)$);
\node[font=\tiny,text=cAcc!80!black,rotate=90,fill=white,inner sep=1.2pt]
  at ($(panel.west)+(-1.2mm,0)$) {AoI work ends here};

\node[font=\tiny,text=cSens!85!black,below=1.3mm of sn.south]
  {$\Dage_i \sim \mathcal{U}(d_i,\,d_i{+}P_i)$};
\node[font=\tiny,text=cRule!85!black,below=2.0mm of panel.south,xshift=4mm]
  {$\Dec=\displaystyle\min_{T\in R}\max_{i\in T}\Dage_i$};

\end{tikzpicture}}
\caption{The command and control chain considered. Existing AoI work
characterises the ages of the individual sources arriving at the fusion node.
This paper characterises what the decision rule downstream of it makes of
them.}
\label{fig:arch}
\end{figure}

Consider a C2 chain fed by $n$ sensing sources, as shown in Fig. \ref{fig:arch}. Source $i$ produces observations of the surveilled state and delivers them to a fusion node. We characterise each source by its stationary age law rather than by a queueing model: this decouples the result from network details and makes heterogeneous sources, such as a rotating radar, a cooperative transponder and a satellite revisit, composable within one formulation. Table \ref{tab:notation} summarises the notation.

\begin{table}[htbp]
\caption{Notation}
\begin{center}
\renewcommand{\arraystretch}{1.1}
\begin{tabular}{@{}ll@{}}
\toprule
\textbf{Symbol} & \textbf{Meaning} \\
\midrule
$n$ & number of sensing sources \\
$P_i$ & sampling period of source $i$ \\
$d_i$ & end-to-end chain delay of source $i$ \\
$\varphi_i$ & sampling phase of source $i$, in $[0,P_i)$ \\
$\lambda_i$ & generation rate of an event-driven source \\
$\delta_i(g)$ & chain delay of the observation generated at $g$ \\
$U_i(t)$ & freshest generation timestamp delivered by $t$ \\
$\Dage_i(t)$ & age of source $i$ at time $t$ \\
$F_i$, $F_R$ & stationary distribution of $\Dage_i$, of $\Dec$ \\
$R$ & decision rule; $T$ a minimal term of $R$ \\
$k$ & corroboration level of a $k$-of-$n$ rule \\
$\Dec(t)$ & decision age under rule $R$ \\
$\Dk$ & $k$-th order statistic of the source ages \\
$\pi_i$ & dominance share of source $i$ \\
$H_n$ & $n$-th harmonic number \\
$\tau_c$ & critical age of the threat class \\
$\varepsilon$ & tolerated violation probability \\
\bottomrule
\end{tabular}
\label{tab:notation}
\end{center}
\end{table}

\subsection{Age of a Source}

Let $U_i(t)$ denote the \emph{generation} timestamp of the freshest observation from source $i$ that has been delivered by time $t$. The age of source $i$ is
\begin{equation}
\Dage_i(t) = t - U_i(t),
\qquad
U_i(t) = \max\{\,g : g + \delta_i(g) \le t\,\},
\label{eq:age}
\end{equation}
where $\delta_i(g)$ is the chain delay experienced by the observation generated at $g$. Two properties follow from \eqref{eq:age} and matter in implementation. First, $U_i$ is a \emph{maximum} over received generation timestamps and not the timestamp of the last arrival: a packet that arrives late but was generated earlier than one already held does not refresh the age, and its delivery consumes capacity for no freshness gain. Second, $\Dage_i$ never falls to zero; on delivery it falls to the chain delay of that observation, because the observation was born before it arrived.

For a source sampling periodically with period $P_i$ and constant chain delay $d_i$, observations are generated at $\varphi_i + mP_i$ for integer $m$, where $\varphi_i \in [0,P_i)$ is the \emph{phase} of the source, that is, the offset of its sampling instants from an arbitrary time origin. The phase matters more than its modest role in \eqref{eq:periodicU} suggests: it is not a design parameter but an accident of when each sensor was switched on, it is neither known nor controlled in a fielded system, and the independence of the source ages assumed below is in the first instance an assumption about phases. With that,
\begin{equation}
U_i(t) = \varphi_i + P_i \left\lfloor \frac{t - d_i - \varphi_i}{P_i} \right\rfloor .
\label{eq:periodicU}
\end{equation}
Observed at an instant uniformly distributed over a period, the stationary age is
\begin{equation}
\Dage_i \sim \mathcal{U}(d_i,\; d_i + P_i),
\qquad
F_i(x) = \left[\frac{x-d_i}{P_i}\right]_0^1 ,
\label{eq:periodiclaw}
\end{equation}
with $[\cdot]_0^1$ denoting clipping to the unit interval, and $\E[\Dage_i] = d_i + P_i/2$. For an event-driven source with Poisson generation of rate $\lambda_i$ and constant delay $d_i$, the residual-age property of the Poisson process gives $\Dage_i - d_i$ exponentially distributed with rate $\lambda_i$, and $\E[\Dage_i] = d_i + 1/\lambda_i$.

\begin{remark}
The decomposition $\E[\Dage_i]=d_i+P_i/2$ already separates two contributions that are routinely conflated: the \emph{chain delay} $d_i$, which is what network engineering optimises, and the \emph{sampling gap} $P_i/2$, which it does not touch. In the scenario of Section \ref{sec:results} the second term exceeds the first by a factor between $2.4$ and $5.2$ for every source.
\end{remark}

\subsection{The Decision Rule}

The alarm condition of an operational system is a monotone Boolean function of which sources currently support the hypothesis. Written in disjunctive normal form, a rule $R$ is a set of minimal terms $T \subseteq \{1,\dots,n\}$, and the rule is satisfied when all sources of some term have contributed. A conjunctive rule, such as ``radar and imagery and AIS-absent'', has a single term; a disjunctive rule, ``any sensor'', has $n$ singleton terms; a $k$-of-$n$ rule has $\binom{n}{k}$ terms of size $k$.

\subsection{The Decision Age}

\begin{proposition}[Decision age]\label{prop:decage}
Let $R$ be a monotone decision rule with minimal terms $T$. The age of the evidence on which the decision is taken at time $t$ is
\begin{equation}
\Dec(t) \;=\; \min_{T \in R} \; \max_{i \in T} \; \Dage_i(t).
\label{eq:decage}
\end{equation}
\end{proposition}

\begin{IEEEproof}
A term is satisfied only once every source it contains has contributed, so the evidence supporting that term is as old as its oldest member, which gives the inner maximum. The rule is satisfied as soon as any term is, so the binding term is the one whose evidence set completes earliest, which gives the outer minimum.
\end{IEEEproof}

For a $k$-of-$n$ rule, \eqref{eq:decage} reduces to the $k$-th order statistic $\Dk$ of the source ages: the rule is satisfied once $k$ sources have reported, and the $k$-th most recent report is the one that completes the evidence. The disjunctive and conjunctive rules are the extreme cases $k=1$, a minimum, and $k=n$, a maximum.

\subsection{Assumptions}

\begin{assumption}[Independence]\label{as:indep}
The stationary ages of distinct sources are mutually independent.
\end{assumption}

Assumption \ref{as:indep} holds when the sampling processes are independent, as they are for physically distinct sensors with uncontrolled phases. It fails in two identifiable cases, and both are worth stating because both are avoidable by design.

First, sources sharing a transmission link or a processing node are coupled through the common queue. We do not treat that case; it is the natural extension of this work.

Second, and less obviously, \emph{periodic sources with identical periods and fixed phases are not independent}: their age difference is the constant $\varphi_j-\varphi_i$, so their ages are perfectly rank-correlated and the order statistics of a single realisation depart from the law of Assumption \ref{as:indep}. The results below are recovered in expectation over phases, and sources with incommensurate periods self-average within a single realisation by equidistribution of the relative phases. Section \ref{sec:results} quantifies both regimes.

\begin{remark}[Design consequence]
Sensors that are to corroborate one another should not be operated at identical synchronised periods: doing so locks their relative ages and removes the averaging that makes the decision age predictable.
\end{remark}

\section{The Decision Age under Fusion Rules}\label{sec:analysis}

\subsection{Distribution}

Under Assumption \ref{as:indep}, the distribution of the decision age under a
$k$-of-$n$ rule follows directly: the rule is satisfied by time $x$ exactly
when at least $k$ of the source ages are below $x$, so
\begin{equation}
\PR(\Dk \le x) \;=\; \PR\!\big(\,\big|\{i : \Dage_i \le x\}\big| \ge k\,\big),
\label{eq:pbdist}
\end{equation}
which is the upper tail of a Poisson-binomial distribution with success probabilities $p_i(x) = F_i(x)$. In particular
\begin{equation}
F_{\wedge}(x) = \prod_{i=1}^{n} F_i(x),
\qquad
F_{\vee}(x) = 1 - \prod_{i=1}^{n}\big(1 - F_i(x)\big)
\label{eq:andor}
\end{equation}
for the conjunctive and disjunctive rules respectively, and for a general monotone rule by inclusion and exclusion over the minimal terms. Equation \eqref{eq:pbdist} is evaluated in $O(n^2)$ by the usual dynamic programme, so no enumeration of the $2^n$ support is required. Writing $F_R$ for the distribution function of $\Dec$ under rule $R$, the mean follows from the tail integral,
\begin{equation}
\E[\Dec] \;=\; \ell \;+\; \int_{\ell}^{h}\!\big(1 - F_{R}(x)\big)\,dx,
\label{eq:mean}
\end{equation}
with $\ell = \min_i d_i$ and $h = \max_i (d_i + P_i)$.

\subsection{Homogeneous Periodic Sources}

\begin{theorem}[$k$-of-$n$ decision age]\label{th:korder}
Let $n$ periodic sources share period $P$ and chain delay $d$, with independent phases. Then $(\Dk - d)/P$ follows a Beta distribution with parameters $k$ and $n-k+1$, and
\begin{equation}
\E\big[\Dk\big] \;=\; d \;+\; P\,\frac{k}{n+1},
\label{eq:theorem1}
\end{equation}
\begin{equation}
\mathrm{Var}\big[\Dk\big] \;=\; P^{2}\,\frac{k\,(n-k+1)}{(n+1)^{2}(n+2)} .
\label{eq:theorem1var}
\end{equation}
\end{theorem}

\begin{IEEEproof}
By \eqref{eq:periodiclaw} the normalised ages $(\Dage_i-d)/P$ are independent and uniform on the unit interval. The $k$-th order statistic of $n$ such variables is Beta distributed with parameters $k$ and $n-k+1$, with mean $k/(n+1)$ and the stated variance. Rescaling to the interval $[d,\,d+P]$ gives \eqref{eq:theorem1} and \eqref{eq:theorem1var}.
\end{IEEEproof}

An elementary reading of \eqref{eq:theorem1} is the following: $n$ points dropped uniformly on a segment cut it into $n+1$ sub-intervals of equal expected length, so the $k$-th point sits at the end of the $k$-th sub-interval and therefore at expected position $k/(n+1)$.

Two special cases of \eqref{eq:theorem1} carry the structural message of the paper. The disjunctive rule gives $\E[\Dage_{(1)}] = d + P/(n+1)$, which tends to the chain delay $d$ as sources are added; the conjunctive rule gives $\E[\Dage_{(n)}] = d + Pn/(n+1)$, which tends to $d+P$. Adding sources under a disjunctive rule drives the decision age towards the \emph{floor} of the age sawtooth; adding them under a conjunctive rule drives it towards the \emph{peak}. The same investment in sensing yields opposite outcomes depending on a line of decision logic, and the dependence on the corroboration level is linear.

\subsection{Event-Driven Sources}

\begin{proposition}[Conjunctive rule, Poisson sources]\label{th:poisson}
Let $n$ event-driven sources have independent Poisson generation of common rate $\lambda$ and common delay $d$. Then
\begin{equation}
\E\big[\Dage_{(n)}\big] \;=\; d \;+\; \frac{H_n}{\lambda},
\qquad
H_n = \sum_{j=1}^{n}\frac{1}{j}.
\label{eq:poisson}
\end{equation}
\end{proposition}

\begin{IEEEproof}
The normalised ages are independent and exponentially distributed. The expectation of their maximum follows by inclusion and exclusion, and for the identically distributed case the resulting alternating sum collapses to $H_n/\lambda$.
\end{IEEEproof}

We state \eqref{eq:poisson} as a proposition rather than a theorem, and deliberately so. The expected maximum of independent exponential variables is a classical order-statistic result, and a harmonic-sum age expression of the same structure is already reported in \cite{bayram2024} for a threshold structure on a gossip network. What is new here is neither the sum nor its derivation, but the object it applies to: the age of a decision taken under a corroboration rule, measured in elapsed time against an operational tolerance.

The consequence, however, is not classical. Comparing \eqref{eq:theorem1} with \eqref{eq:poisson}, the cost of corroboration grows \emph{linearly and saturates} for periodic sampling, and grows \emph{logarithmically without bound} for Poisson sampling. The sampling process, and not only its rate, determines the price of requiring corroboration.

\subsection{Dominance}

Two facts bound the problem. For a conjunctive rule $\E[\Dage_{(n)}] \ge \max_i \E[\Dage_i]$, so the decision is never fresher than its worst required source; and if a source $j$ satisfies $d_j > d_i + P_i$ for every $i \ne j$, then $\Dage_{(n)} = \Dage_j$ almost surely, so that source alone sets the freshness. That is the extreme case. The operationally useful quantity is the \emph{dominance share} of each source, defined as the fraction of decision instants at which source $i$ supplies the binding $k$-th order statistic,
\begin{equation}
\pi_i \;=\; \PR\big(\,\Dage_i = \Dk\,\big).
\label{eq:dominance}
\end{equation}
The resulting vector answers a question that system engineers ask and currently cannot answer: which sensor is actually setting the freshness of the decisions, and therefore where does improvement change the outcome.

\subsection{The Freshness Requirement}

Let $\tau_c$ be the \emph{critical age} of a threat class, obtained from its kinematics as the reaction distance set by the concept of operations divided by the maximum threat speed. The system requirement becomes
\begin{equation}
\PR\big(\Dec > \tau_c\big) \;\le\; \varepsilon,
\label{eq:requirement}
\end{equation}
which, for periodic sources under a conjunctive rule, admits the closed form
\begin{equation}
\PR\big(\Dage_{(n)} > \tau_c\big) \;=\; 1 - \prod_{i=1}^{n}
\left[\frac{\tau_c - d_i}{P_i}\right]_0^1 .
\label{eq:reqclosed}
\end{equation}

Equation \eqref{eq:requirement} is agnostic to the technology: it does not presuppose whether the requirement is met with bandwidth, with scheduling, with additional sensors or with a change of decision rule. It is verifiable by instrumentation, and it is traceable to the operation through $\tau_c$.

\section{Measuring the Decision Age}\label{sec:measure}

The formulation above is of practical use only if the quantity it describes can be measured in a fielded system. Three requirements apply.

\paragraph{Generation timestamps} The age in \eqref{eq:age} is defined from the instant of \emph{generation}, which must travel with the observation. It must mark the instant at which the physical state was sampled and not the instant at which processing completed: timestamping at the output of a detector hides the processing latency inside the apparent freshness, and the resulting measurement is systematically optimistic.

\paragraph{Clock discipline} Generation timestamps from different sources are compared, so clock offset translates directly into age error and must be budgeted. Where satellite navigation is denied, the drift must be monitored, because it corrupts the metric silently.

\paragraph{An unbiased estimator} Sampling the age sawtooth uniformly and averaging is biased. The area under the sawtooth is instead accumulated exactly and event by event, with constant state per source: an accumulated area $A$, an accumulated elapsed time $E$, the freshest generation timestamp $U$ held so far, and the instant $a_{\mathrm{prev}}$ of the previous delivery. On delivery at time $a$ of an observation generated at $g$,
\begin{align}
A &\leftarrow A + \tfrac{1}{2}\Big[(a-U)^2 - (a_{\mathrm{prev}}-U)^2\Big],
\label{eq:estimator}\\
E &\leftarrow E + (a - a_{\mathrm{prev}}), \quad
U \leftarrow \max(U, g), \quad a_{\mathrm{prev}} \leftarrow a,
\nonumber
\end{align}
and the time-average age is $A/E$. The same pass yields the peak-age sample $a-U$ and the per-packet delay $a-g$, so the three quantities are obtained together.

\paragraph{Where to sample the distribution} For the requirement \eqref{eq:requirement} the relevant distribution is not that of the decision age over time, but that of the decision age \emph{at the instants at which the decision rule is evaluated}. The age vector is therefore recorded at each rule evaluation, which is the query-based sampling discussed in \cite{yates2021}.

\section{Numerical Results}\label{sec:results}

All analytical expressions of Section \ref{sec:analysis} are validated against
discrete-event simulation. The simulator draws an independent phase for every
source and records the age vector at the instants at which the decision rule is
evaluated, which is the query-based sampling argued for in Section
\ref{sec:measure}. Seeds are fixed, so every figure and every number below is
reproducible; the code and the parameter set are released.

\subsection{Scenario}\label{sec:results-scenario}

We instantiate the formulation on a port-approach surveillance chain with
$n=6$ heterogeneous sources, summarised in Table \ref{tab:scenario}. No figure
in this paper comes from an operational system.

The last column of Table \ref{tab:scenario} distinguishes two kinds of
quantity, because they are not interchangeable. Two rows are fixed by public
specification (S). The AIS period is the Class~A autonomous reporting interval
for a vessel under way at moderate speed \cite{itu1371}. The radar period is
one antenna revolution, which is the natural time quantum of a scanning
sensor: the international performance standard defines detection itself per
scan, requiring the target in at least 8 of 10 scans, while leaving the
rotation rate to the installation \cite{imo192}; the value used, 16.2~rpm,
lies in the slow part of the 15 to 45~rpm published for a representative
coastal surveillance radar \cite{terma1002}.

The remaining four rows are design choices (D), and the electro-optical and
infrared rows are worth a comment because the distinction is easy to get
wrong. A surveillance camera publishes a frame rate of tens of hertz
\cite{flirfh}, which is three orders of magnitude faster than the period used
here. That frame rate is not the quantity this model needs. What determines
when a fresh \emph{detection} becomes available to the fusion node is the
cadence of the downstream processing chain, and no public standard or
datasheet fixes it; the same is true of the revisit interval of an unmanned
patrol orbit and of the processing cadence of a hydrophone array. These follow
from the concept of operations, and we state them as such rather than attach a
citation that would not support them. The observation generalises: taking a
sensor's nominal rate for its effective one is a modelling error that makes a
system look an order of magnitude fresher than it is.

The distinction matters less than it appears, because no conclusion of this
paper depends on the specific values. What the analysis requires is only that
the cadences be heterogeneous and the rule be strict, and Section
\ref{sec:results-sensitivity} settles that point rather than asserting it: it
varies the chain by an order of magnitude, adds a satellite hop, flattens the
suite to a homogeneous cadence and relaxes the rule, and reports the regime in
which the conclusion holds and the regime in which it does not.

Two choices in Table \ref{tab:scenario} are deliberate. First, the sampling
periods are mutually incommensurate. Sources with identical periods violate
Assumption \ref{as:indep}, as quantified in Section
\ref{sec:results-commensurability}. Second, the chain delays are small relative
to the sampling periods, which is the realistic regime: across all six sources
the sampling gap $P_i/2$ exceeds the chain delay $d_i$ by a factor between
\ResRatioMin{} and \ResRatioMax{}, as anticipated in Remark 2.

\begin{table}[htbp]
\caption{Port-approach surveillance suite. The last column states what each
sampling period is: S denotes a value fixed by public specification, D a design
choice of the concept of operations. Section \ref{sec:results-sensitivity}
shows that no conclusion depends on the specific values.}
\begin{center}
\footnotesize
\renewcommand{\arraystretch}{1.15}
\begin{tabular}{@{}lrrcl@{}}
\toprule
\textbf{Source} & $P_i$ [s] & $d_i$ [s] & & \textbf{What $P_i$ is} \\
\midrule
Coastal radar & 3.7 & 0.50 & S & Antenna revolution \cite{terma1002} \\
AIS           & 10.0 & 1.00 & S & Class A reporting \cite{itu1371} \\
EO camera     & 1.5 & 0.30 & D & EO detection cadence \\
IR camera     & 1.7 & 0.35 & D & IR detection cadence \\
UAS feed      & 8.3 & 0.80 & D & Patrol orbit revisit \\
Acoustic      & 2.3 & 0.40 & D & Hydrophone processing \\
\bottomrule
\end{tabular}
\label{tab:scenario}
\end{center}
\end{table}

The critical age follows from threat kinematics. For a fast patrol craft at
approximately 30 knots and a reaction perimeter of 100 metres,
$\tau_c = \ResTauC$ s. This value is used throughout unless stated otherwise.

\subsection{Model Validation}\label{sec:results-validation}

\begin{figure}[htbp]
\centerline{\includegraphics[width=\columnwidth]{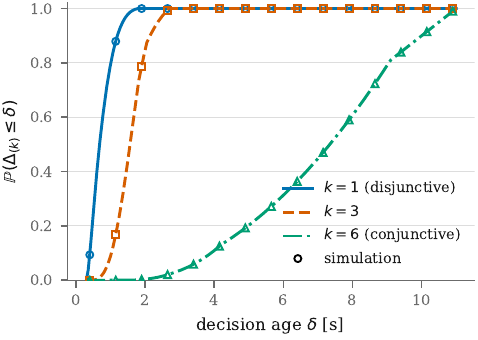}}
\caption{Analytical decision-age distribution (lines) against discrete-event
simulation (markers), for the disjunctive, intermediate and conjunctive rules.
Maximum deviation \ResMaxCdfError{} in probability.}
\label{fig:validation}
\end{figure}

Fig. \ref{fig:validation} compares the analytical distribution of
\eqref{eq:pbdist} against simulation for three rules: disjunctive, an
intermediate $k=3$, and conjunctive. Lines are the model and markers the
empirical distribution obtained from \ResNDecisions{} decision instants.

The maximum absolute deviation is \ResMaxCdfError{} in probability, and
\ResMaxMeanError{} s in the mean decision age across all six values of $k$.
The closed forms are therefore exact to within simulation noise, and this
agreement is the empirical basis for every result that follows.

\subsection{The Mechanism}\label{sec:results-sawtooth}

Fig. \ref{fig:sawtooth}, shown in Section \ref{sec:intro-motivation} before any
of the analysis, is generated from this scenario: a 42-second window with three
of the six sources and the decision age of the conjunctive rule superimposed.
It is worth returning to it now that the quantities have been defined.

The decision age does not track the average of the source ages, nor the
freshest: it tracks the \emph{slowest required} source. The radar and the
camera refresh continuously throughout the window, and
\ResNFreeSources{} of the three contribute nothing at all to the freshness of
the decision, because the conjunction is always waiting for the third. The
decision is stale \ResStaleFrac{}\% of the time while neither fast sensor is
ever stale. This is the asymmetry that the dominance shares of Section
\ref{sec:results-dominance} quantify across the full suite: freshness is set
by a minority of the sources, and the rest of the sensing budget buys
detection performance without buying currency.

\subsection{Decision Age versus Corroboration Level}\label{sec:results-corroboration}

\begin{figure*}[t]
\centerline{\includegraphics[width=\textwidth]{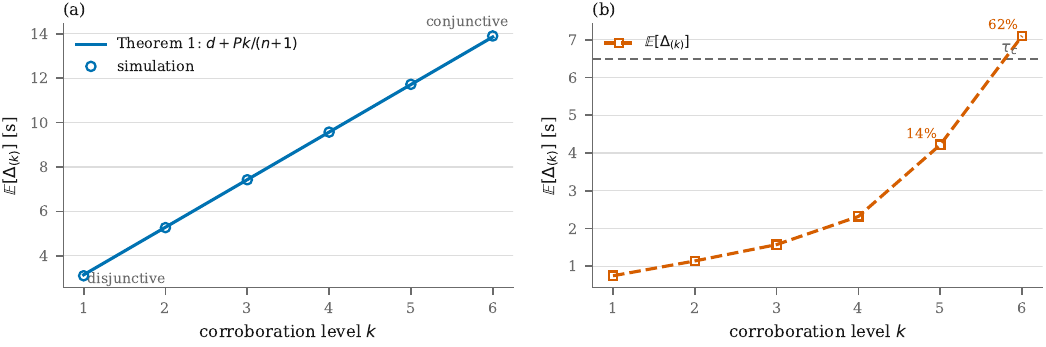}}
\caption{Mean decision age against corroboration level. (a) Homogeneous
sources, $n=6$: Theorem \ref{th:korder} against simulation; the dependence is
exactly linear. (b) Heterogeneous port suite of Table \ref{tab:scenario}: the
dependence becomes convex, and the annotations give the probability of
violating the freshness requirement.}
\label{fig:corroboration}
\end{figure*}

Fig. \ref{fig:corroboration}(a) validates Theorem \ref{th:korder} on a
homogeneous suite. The analytical line and the simulated points agree to
\ResThmVsSim{} s, and the closed form matches direct numerical evaluation of
\eqref{eq:mean} to \ResThmVsModel{}. The dependence on $k$ is exactly linear,
as \eqref{eq:theorem1} requires.

Fig. \ref{fig:corroboration}(b) repeats the computation on the heterogeneous
suite of Table \ref{tab:scenario}, which is the operationally relevant case.
The mean decision age rises from \ResAgeKone{} s under a disjunctive rule to
\ResAgeKsix{} s under a conjunctive one, a factor of \ResAgeRatio{}. The
dependence is no longer linear: it is convex, because tightening the rule
progressively recruits the slower sources.

The consequence for the freshness requirement is sharper than the means
suggest. Against $\tau_c = \ResTauC$ s the requirement is met without
exception for $k \le 4$; at $k=5$ it is violated \ResViolKfive{}\% of the
time, and at $k=6$ \ResViolKsix{}\% of the time. The transition from a
compliant system to a non-compliant one is produced entirely by the decision
rule, with no change to any sensor or to any link.

\subsection{Where the Decision Age Comes From}\label{sec:results-decomposition}

It is natural to ask how much of the decision age is attributable to sampling,
how much to the communication chain, and how much to the decision rule itself.
The question is not trivial to answer, because the decision age is a maximum
and not a sum: attributing contributions by switching factors on one at a time
gives an answer that depends on the order chosen.

\begin{figure}[htbp]
\centerline{\includegraphics[width=\columnwidth]{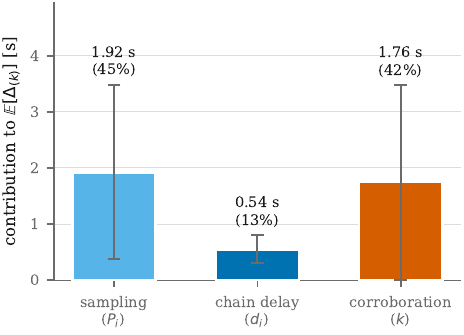}}
\caption{Shapley attribution of the mean decision age over three factors, at
$k=5$. Error bars give the range over all orderings of the factors, not
statistical uncertainty. The three contributions sum exactly to the total of
\ResDecTotal{} s.}
\label{fig:decomposition}
\end{figure}

We therefore compute the exact Shapley attribution over the three factors
$\{$sampling, chain delay, corroboration$\}$, averaging over all $3!$
orderings. The three contributions then sum exactly to the total, and the
attribution is order-independent by construction. Fig. \ref{fig:decomposition}
reports the Shapley values with error bars giving the range over orderings.

For the baseline suite at $k=5$, the mean decision age of \ResDecTotal{} s
decomposes into \ResShareSampling{}\% from sampling, \ResShareDelay{}\% from
the communication chain and \ResShareCorrob{}\% from the corroboration
requirement. The corroboration requirement contributes \ResCorrobOverDelay{}
times as much as the chain delay.

This is the practical point of the paper for a system designer. The
communication chain is the component that receives engineering attention and
procurement budget, and it accounts for roughly an eighth of the staleness of
the decision. The two quantities that dominate, sampling cadence and the
corroboration level written into the alarm rule, are not usually regarded as
freshness parameters at all.

\subsection{Sensitivity}\label{sec:results-sensitivity}

The attribution above holds in a regime, not universally, and Table
\ref{tab:sensitivity} delimits it.

Degrading the communication chain by an order of magnitude does not overturn
the conclusion: the corroboration share falls only to \ResShareCorrobTenX{}\%
while the chain rises to roughly half, and the total age nearly triples. This
addresses the most likely objection, that the baseline assumes an unusually
good network.

The conclusion is however genuinely conditional on two properties of the
system. With a homogeneous sensor suite the corroboration share falls to
\ResShareCorrobHomog{}\%, and with a relaxed rule to
\ResShareCorrobKthree{}\%. We therefore state the result in the form it
actually holds: \emph{the decision rule dominates the decision age when the
source cadences are heterogeneous and the rule is strict}. Both conditions are
characteristic of operational surveillance, where sensors ranging from seconds
to minutes of cadence are combined and corroboration is imposed to contain the
false alarm rate.

\begin{table}[htbp]
\caption{Sensitivity of the attribution}
\begin{center}
\renewcommand{\arraystretch}{1.15}
\begin{tabular}{@{}lrrr@{}}
\toprule
\textbf{Configuration} & \textbf{Total [s]} & \textbf{Chain} & \textbf{Rule} \\
\midrule
Baseline, $k=5$              & 4.22  & 13\% & 42\% \\
Chain delay $\times 10$      & 11.65 & 48\% & 34\% \\
Satellite hop, $+600$ ms     & 4.82  & 24\% & 37\% \\
Homogeneous cadence          & 2.00  & 25\% & 34\% \\
Relaxed rule, $k=3$          & 1.57  & 24\% & 27\% \\
\bottomrule
\end{tabular}
\label{tab:sensitivity}
\end{center}
\end{table}

\subsection{Dominance}\label{sec:results-dominance}

\begin{figure}[htbp]
\centerline{\includegraphics[width=\columnwidth]{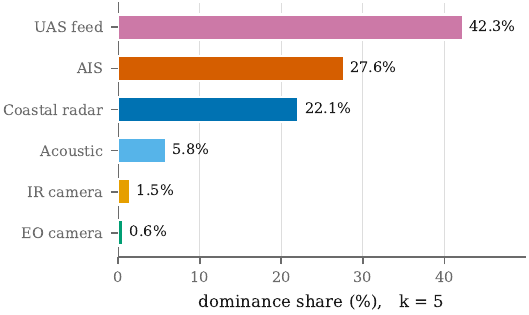}}
\caption{Dominance shares at $k=5$: the fraction of decision instants at which
each source supplies the binding order statistic, and therefore sets the
freshness of the decision.}
\label{fig:dominance}
\end{figure}

Fig. \ref{fig:dominance} reports the dominance shares of \eqref{eq:dominance}:
the fraction of decision instants at which each source supplies the binding
order statistic.

The distribution is strongly unequal. The \ResDomTopName{} alone governs
\ResDomTop{}\% of decisions and the two leading sources together govern
\ResDomTwo{}\%, while the two cameras combined govern \ResDomCameras{}\%.
The engineering reading is immediate and is not available from any
conventional metric: effort spent on the electro-optical and infrared chains
cannot measurably improve decision freshness, however much it improves those
sensors in isolation, whereas the revisit interval of the unmanned platform
and the reporting cadence of the cooperative transponder set it almost
entirely.

\subsection{Cost of Corroboration under Different Sampling Processes}
\label{sec:results-scaling}

\begin{figure}[htbp]
\centerline{\includegraphics[width=\columnwidth]{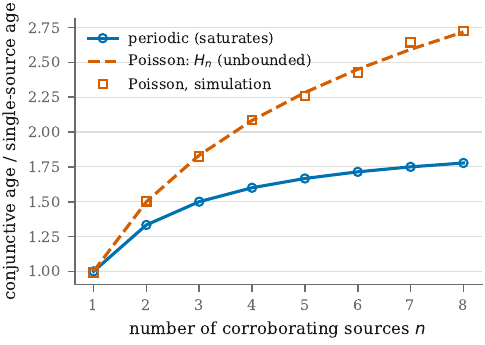}}
\caption{Growth of the conjunctive decision age with the number of
corroborating sources, normalised by the single-source mean age. Periodic
sampling saturates; Poisson sampling grows as $H_n$ without bound.}
\label{fig:scaling}
\end{figure}

Fig. \ref{fig:scaling} compares the growth of the conjunctive decision age
with the number of corroborating sources, for periodic and for event-driven
sampling, normalised in both cases by the mean age of a single source so that
the two curves start together.

The contrast predicted by Theorem \ref{th:korder} and Proposition \ref{th:poisson} is
clear. Periodic sampling saturates: at $n=8$ the conjunctive age is
\ResPeriodicEight{} times the single-source age, and it cannot exceed twice it
however many sources are added, since the age is bounded by $d+P$. Poisson
sampling has no such bound and grows as the harmonic numbers, reaching
\ResPoissonEight{} times at $n=8$ and continuing to grow. Simulation matches
the analytical harmonic expression to \ResPoissonError{}.

The design consequence is that the \emph{type} of sampling process, not only
its rate, determines the price of requiring corroboration. A suite of
event-driven sources pays an unbounded penalty for corroboration that a suite
of cadenced sources does not.

\subsection{Degraded Operation}\label{sec:results-degraded}

The sensitivity analysis above worsens the chain but leaves it perfect and
permanent. Operational systems fail in two other ways, and both act on the
decision age in ways the nominal analysis does not anticipate.

\begin{figure}[t]
\centerline{\includegraphics[width=\columnwidth]{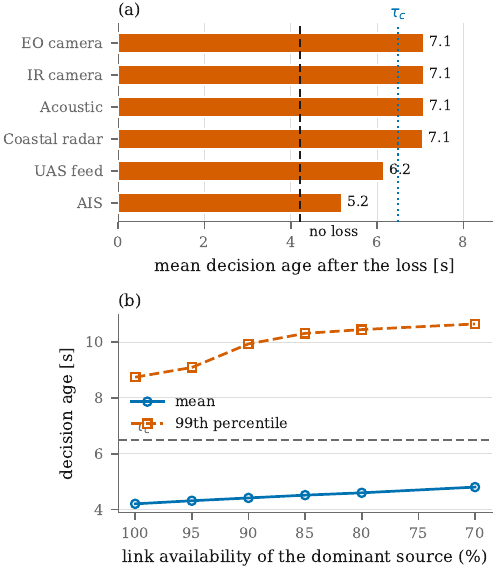}}
\caption{Degraded operation. (a) Mean decision age after the loss of each
source, against nominal operation and the tolerance $\tau_c$; the rule still
demands $k=5$ from the survivors. Losing a fast sensor is worse than losing a
dominant one, because it forces the slow sources into the required set.
(b) Sweep of the link availability of the dominant source: the mean is almost
insensitive to intermittency while the tail is not.}
\label{fig:degraded}
\end{figure}

\paragraph{Loss of a source} When a sensor stops reporting, the rule does not
relax itself: it still demands $k$ corroborations from whatever remains. For
$k=n$ the rule becomes \emph{unsatisfiable} and the decision age is not large
but undefined, which is a failure mode worth stating explicitly in a design
document. For $k<n$ the decision survives and ages, and Fig.
\ref{fig:degraded}(a) shows by how much.

The result is counter-intuitive, and it refines rather than contradicts the
dominance analysis of Section \ref{sec:results-dominance}. Losing the
\ResLossWorstName{} takes the mean decision age to \ResLossWorstAge{}~s and
the violation probability to \ResLossWorstViol{}\%, whereas losing the
\ResLossBestName{} takes them only to \ResLossBestAge{}~s and
\ResLossBestViol{}\% --- although the \ResLossWorstName{} governs
\ResDomCameras{}\% of decisions in nominal operation.

The explanation is structural: with one source gone, a $k=5$ rule over five
survivors is conjunctive, so losing a \emph{fast} source forces the slow ones
into the mandatory set, while losing the slowest removes the binding
constraint. The consequence is easy to get wrong. \textbf{The dominance vector
says where investment improves freshness, not what may safely be removed.} A
sensor with a negligible dominance share can be the one whose loss breaks the
requirement, because its role is to spare the rule from the slower sources.

\paragraph{An intermittent link} Fig. \ref{fig:degraded}(b) sweeps the link
availability of the dominant source down to $70$\%, with outages of eight
seconds on average. The mean decision age moves from \ResAvailMeanLo{} to
\ResAvailMeanHi{}~s, which a specification written on the mean would judge
marginal; over the same sweep the $99$th percentile moves from
\ResAvailTailLo{} to \ResAvailTailHi{}~s and the violation probability from
\ResAvailViolLo{}\% to \ResAvailViolHi{}\%. The argument is for the form of
the requirement, not for a number: written on the mean it is nearly blind to
intermittency, which is the failure mode a tactical link actually exhibits,
and written as \eqref{eq:requirement} it is not.

\subsection{When the Independence Assumption Fails}
\label{sec:results-commensurability}

\begin{figure}[htbp]
\centerline{\includegraphics[width=\columnwidth]{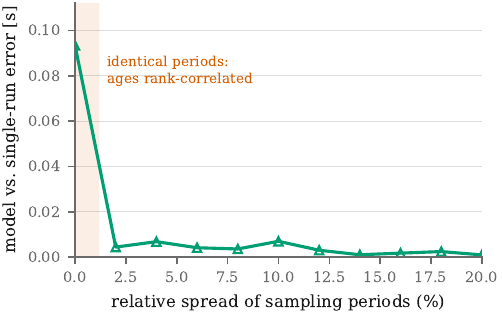}}
\caption{Discrepancy between the analytical decision age and a single
simulated realisation, against the relative spread of the sampling periods.
Identical periods break Assumption \ref{as:indep}; a few per cent of mismatch
restores it.}
\label{fig:commensurability}
\end{figure}

Section \ref{sec:model} noted that periodic sources with identical periods and
fixed phases are not independent, because their age differences are constant.
Fig. \ref{fig:commensurability} quantifies the effect. Six sources are given
progressively mismatched periods, from identical to a relative spread of 20\%,
and for each spread the analytical mean decision age is compared against a
\emph{single} simulated realisation, which is what a fielded system provides.

With identical periods the discrepancy is \ResCommErrZero{} s. A spread of a
few per cent is sufficient to remove it, and at 20\% the discrepancy is
\ResCommErrSpread{} s, two orders of magnitude smaller. The relative phases of
incommensurate periods equidistribute, and a single realisation then
self-averages.

This yields the design guidance of Remark 3 in quantitative form. Sensors
intended to corroborate one another should not be operated at identical
synchronised periods: doing so locks their relative ages, removes the
averaging on which the closed forms depend, and makes the decision age of the
deployed system depend on an arbitrary and uncontrolled phase alignment.

\section{Implications for C2 System Design}\label{sec:implications}

This section is written for the engineer who specifies, procures or accepts a
multi-sensor C2 system, and each item below is something that can be acted on
without adopting the rest of the formulation.

\subsection{Writing a Freshness Requirement That Can Be Verified}

A requirement of the form ``end-to-end latency shall not exceed $X$~ms'' has
three defects: it prescribes a solution, it constrains a component rather than
the product, and it is silent about the decision that the system exists to
take. Equation \eqref{eq:requirement} replaces it with a statement that has
none of those defects:

\begin{quote}\itshape
For threat class $c$ and alarm rule $R$, the probability that the evidence on
which the alarm is raised is older than $\tau_c$ shall not exceed
$\varepsilon$.
\end{quote}

\noindent It is agnostic to the technology, so a bidder may meet it with
bandwidth, with scheduling, with sampling cadence, with a different fusion rule
or with any combination. It is traceable to the operation, because $\tau_c$ is
the reaction distance of the concept of operations divided by the maximum
threat speed. And it is verifiable in the field with the instrumentation of
Section \ref{sec:measure}, which costs three counters per source. For
procurement and acceptance this is, in our judgement, the most immediately
applicable consequence of the formulation.

\subsection{Spending the Sensing Budget Where It Changes the Outcome}

The dominance shares of \eqref{eq:dominance} answer a question that programme
offices ask and currently cannot answer: which sensor is setting the freshness
of the decisions. In the suite of Table \ref{tab:scenario} the answer is
strongly unequal, with \ResDomTwo{}\% of decisions governed by two sources and
\ResDomCameras{}\% by the two cameras together. Shortening the processing
chain of the electro-optical sensor, or upgrading its focal plane, cannot
measurably improve decision freshness in that configuration; shortening the
revisit interval of the unmanned platform can. The dominance vector is cheap
to compute, requires only the cadences and delays that an integrator already
knows, and should be produced before the sensing budget is allocated rather
than after.

\subsection{Pricing the Corroboration Level}

Corroboration levels are chosen on detection grounds, to contain the false
alarm rate, by engineers who are not considering time, and the choice is
recorded in a configuration file rather than in a requirement. Theorem
\ref{th:korder} prices it. In the scenario of Section \ref{sec:results},
tightening the rule from $k=4$ to $k=6$ takes the system from meeting the
freshness requirement without exception to violating it \ResViolKsix{}\% of
the time, with no change to any sensor and no change to any link. The
corroboration requirement contributes \ResCorrobOverDelay{} times as much to
the staleness of the decision as the whole communication chain.

The practical recommendation follows: the corroboration level belongs in the
requirement baseline, under configuration control, alongside the false alarm
rate it was chosen to contain. A parameter that can move a compliant system to
a non-compliant one is not a tuning knob.

\subsection{Showing the Operator What They Are Looking At}

An operator sees a track symbol with no indication of its age, and a symbol
two seconds old is indistinguishable from one four minutes old. Rendering the
age normalised by the critical age, $\Dec/\tau_c$, puts the two on different
footings with a single number that needs no explanation: below one the
decision is current, above one it is not. This is a trust-calibration measure
that follows directly from the formulation, costs little to implement, and
does not require the operator to know anything about the analysis.

\subsection{Age as a Condition on Delegated Authority}

Where engagement decisions are delegated to autonomous or semi-autonomous
platforms, the question of how stale the evidence may be for an authorisation
to remain valid is one of safety and accountability rather than of
performance. Conditioning an authorisation on a bound on the decision age is
implementable with the same instrumentation, is auditable after the fact, and
the age recorded at the decision instant is evidence of due diligence. We note
this as a direction rather than a result: the bound itself is a policy
question, not an engineering one, and the formulation offers the quantity to
write the policy against, not the policy.

\section{Conclusion}\label{sec:conclusion}

We have defined the decision age of a monotone fusion rule and shown that, under corroboration, information freshness is governed by a maximum rather than a minimum of source ages, which is the opposite of the behaviour assumed in the multi-source AoI literature. For periodic sources the decision-age distribution is available in closed form, and the mean under a $k$-of-$n$ rule is $d + Pk/(n{+}1)$, moving linearly from the floor of the age sawtooth to its peak as corroboration is tightened. A dominance criterion identifies the source that governs decision freshness, and the formulation turns freshness into a verifiable requirement.

Two extensions follow directly. The first is the correlated case, in which sources share a link or a processing node and the independence assumption is relaxed. The second is experimental validation on an operational multi-sensor surveillance system, using the instrumentation of Section \ref{sec:measure}; this is the subject of ongoing work.


%
%
%
%
\bibliographystyle{IEEEtran}
\bibliography{refs}

\end{document}